\documentclass[prb,
                    preprint
                    ]{revtex4}
\usepackage{epsfig}

\def\e{\begin{equation}}
\def\f{\end{equation}}
\def\ea{\begin{eqnarray}}
\def\fa{\end{eqnarray}}
\def\=#1{\overline{\overline{#1}}}
\def\_#1{{\bf #1}}

\def\.{\cdot}

\def\l#1{\label{eq:#1}}
\def\r#1{(\ref{eq:#1})}

\def\l#1{\label{eq:#1}}
\def\r#1{(\ref{eq:#1})}

\begin{document}

\title{Three-dimensional isotropic perfect lens based on \textit{L}\textit{C}-loaded
transmission lines}

\author{Pekka Alitalo}

\author{Stanislav Maslovski}

\author{Sergei Tretyakov}

\affiliation{Radio Laboratory / SMARAD, Helsinki University of Technology\\
P.O. Box 3000, FI-02015 TKK, Finland\\
{\rm E-mails: pekka.alitalo@hut.fi, stanislav.maslovski@hut.fi,
sergei.tretyakov@hut.fi}}

\date{\today}

\begin{abstract}
An isotropic three-dimentional perfect lens based on cubic meshes
of interconnected transmission lines and bulk loads is proposed.
The lens is formed by a slab of a loaded mesh placed in between
two similar unloaded meshes. The dispersion equations and the
characteristic impedances of the eigenwaves in the meshes are
derived analytically, with an emphasis on generality. This allows
designing of transmission-line meshes with desired dispersion
properties. The required backward-wave mode of operation in the
lens is realized with simple inductive and capacitive loads. An
analytical expression for the transmission through the lens is
derived and the amplification of evanescent waves is demonstrated.
Factors that influence enhancement of evanescent waves in the lens
are studied and the corresponding design criteria are established.
A possible realization of the structure is outlined.
\end{abstract}

\maketitle

\section{Introduction}

In recent literature a lot of attention has been given to systems
that are able to sense electromagnetic near fields (evanescent
waves) and even to ``amplify'' them. The superlens proposed by
Pendry\cite{Pendry} is one of such systems. His superlens is based
on a Veselago medium\cite{Veselago} slab. The real parts of the
permittivity and the permeability of the Veselago slab are both
negative at a certain wavelength. Thus, the eigenwaves in the slab
are {\it backward} waves, i.e. the wave phase and group velocities
are antiparallel. This provides negative refraction and focusing
of waves in a planar slab, as was outlined by
Veselago.\cite{Veselago} However, Pendry discovered that there was
also a possibility to excite surface plasmon-polaritons on the
slab surfaces and, due to that, amplify near fields\cite{Pendry}.
The slab thickness can be of the order of the wavelength, so that
the plasmon-polaritons excited at both sides of the slab are
strongly coupled. Under certain conditions the plasmon-polariton
excited at the back surface of the slab has a much stronger
amplitude than that at the front surface. Such amplification of
evanescent waves is the key principle in subwavelength imaging.

Known experimental realizations of volumetric artificial materials
with negative parameters are highly anisotropic structures that
utilize dense arrays of parallel thin conducting wires and
variations of split-ring resonators\cite{science}. Proposed
isotropic arrangements use orthogonal sets of split
rings\cite{Gay} and also three-dimensional arrays of
wires\cite{iso}. At the same time, there have been achievements in
modeling of Veselago materials (and Pendry lens) with the help of
$LC$-circuits or transmission-line (TL) based
structures.\cite{Eleftheriades,history,Sanada,Grbic1,
Grbic2,Grbic3} These networks do not rely on resonant response
from particular inclusions, and the period of the mesh can be made
very small as compared with the wavelength. These features allow
realization of broadband and low-loss devices, which is extremely
difficult if resonant inclusions are used.

The transmission-line network approach has been successfully
realized in one- and two-dimensional networks, but up to now there
have been doubts if it is possible to design a three-dimensional
(3D) circuit analogy of the Veselago medium. The difficulties
arise from the fact that such a 3D network requires a common
ground connector. Any realization of such a ground will
effectively work as a dense mesh of interconnected conductors that
blocks propagation of the electromagnetic waves practically the
same way as a solid metal does (the structural period must be much
less than the wavelength in order to realize an effectively
uniform artificial material). In this paper we introduce isotropic
three-dimensional transmission-line networks that overcome this
difficulty.

In the TL-based networks that we study,
the electromagnetic energy propagates through TL
sections. The inside of every TL section is effectively
screened from the inside of the other sections and from the outer space. This can be
very naturally imagined with a 3D cubic-cell network of
interconnected coaxial cable segments: The inner conductors of the
segments are soldered at the network nodes; the same is done for
the outer conductors. The whole system appears as a 3D pipe
network where every pipe holds a central conductor and those
conductors are crossing at the node points inside the pipes. A
loaded TL network can be realized now by placing loading elements
inside the ``pipes''. To couple the waves propagating inside the TL sections with
the free-space waves one will have to apply a kind of antenna array
with every antenna feeding a particular TL segment.

When using transmission lines loaded with bulk elements we speak of
waves in the meaning of discrete waves of voltages
and currents defined at the loading positions. Let us note that in
the TL sections {\it as such} the usual, forward waves propagate.
Only because of the loading the discrete voltage and current waves
appear as backward ones when appropriate loading impedances are used.

While completing this manuscript, we learned about another possible
design of a 3D transmission-line analogy of a backward-wave material described in Ref.~12.
That design is based on Kron's formal representation of Maxwell's
equations as an equivalent electric circuit.\cite{Kron} In Ref.~12 only 1D
propagation was studied analytically and 3D properties were
analyzed numerically.

\section{Three-dimensional transmission-line networks}

\begin{figure}[ht]
\centering \epsfig{file=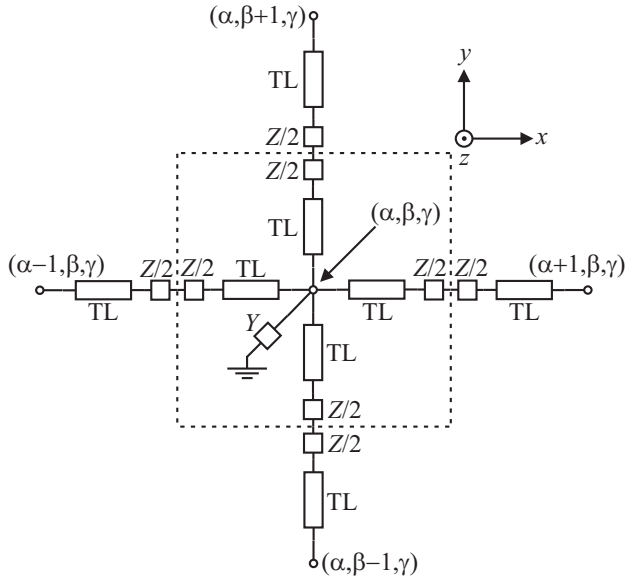,
width=0.5\textwidth} \caption{Unit cell of a 3D backward-wave
transmission line network (enclosed by the dotted line). The
transmission lines and impedances along  the $z$-axis are not
shown. Transmission lines have the characteristic impedance $Z_0$
and length $d/2$ ($d$ is the period of the structure).}
\label{3D_ZY_TL_unit_cell}
\end{figure}

The proposed structure of 3D super-resolution lens consists of two
forward-wave (FW) regions and one backward-wave (BW) region. The
3D forward-wave networks can be realized with simple transmission
lines and the 3D backward-wave network with inductively and
capacitively loaded transmission lines. One unit cell of the BW
network is shown in Fig.~\ref{3D_ZY_TL_unit_cell} (the unit cell
enclosed by the dotted line). In the 3D structure there are
impedances \textit{Z}/2 and transmission lines also along the
$z$-axis (not shown in Fig.~\ref{3D_ZY_TL_unit_cell}). In view of
potential generalizations, the loads are  represented by series
impedances \textit{Z}/2 and shunt admittances \textit{Y}, although
for our particular purpose to realize a backward-wave network, the
loads are simple  capacitances and inductances. The unit cell of
the FW network is the same as in Fig.~\ref{3D_ZY_TL_unit_cell} but
without the series impedances $Z/2$ and shunt admittance $Y$. The
equations that will be derived for these structures can be used in
various implementations, but this paper will concentrate on the
case when $Z=1/j\omega C$ and $Y=1/j\omega L$.

\section{Dispersion in backward-wave and forward-wave networks}

\subsection{Dispersion equations}

First we will derive the dispersion relation for a simplified 3D
BW network, i.e. we will not take into account the transmission
lines. Such approximation is possible at low frequencies. Without
the transmission line segments this derivation is quite simple and
can be done by summing up all the currents that flow to the node
($\alpha,\beta,\gamma$) and equating this sum with the current
flowing to the ground (through admittance $Y$), see
Fig.~\ref{3D_ZY_TL_unit_cell}. The result is \e
\frac{1}{Z}(U_{\alpha+1,\beta,\gamma}+U_{\alpha,\beta+1,\gamma}+U_{\alpha,\beta,
\gamma+1}+U_{\alpha-1,\beta,\gamma}+U_{\alpha,\beta-1,\gamma}+U_{\alpha,\beta,\gamma-1}-6U
_{\alpha,\beta,\gamma})=U_{\alpha,\beta,\gamma}Y \l{simple_1}. \f
We look for a solution of the form
$U_{\alpha,\beta,\gamma}=U_0e^{-j\vec{k}\cdot \vec{r}}$
($\vec{r}=r_x\vec{x_0}+r_y\vec{y_0}+r_z\vec{z_0}$), and if we use
$q_{x}=k_{x} d$, $q_{y}=k_{y} d$ and $q_{z}=k_{z} d$, (1) can be
reduced to \e
\frac{1}{Z}(e^{-jq_{x}}+e^{-jq_{y}}+e^{-jq_{z}}+e^{+jq_{x}}+e^{+jq_{y}}+e^{+jq_{z}}-6)=Y,
 \f
or
\e \cos(q_{x})+\cos(q_{y})+\cos(q_{z})=\frac{ZY}{2}+3. \f

If we now insert $Z=1/j\omega C$ and $Y=1/j\omega L$ we get the
dispersion relation for the \textit{L}\textit{C}-loaded network:
\e \cos(q_{x})+\cos(q_{y})+\cos(q_{z})=-\frac{1}{2\omega^{2}LC}+3.
\l{simple_3} \f

\begin{figure}[ht]
\centering \epsfig{file=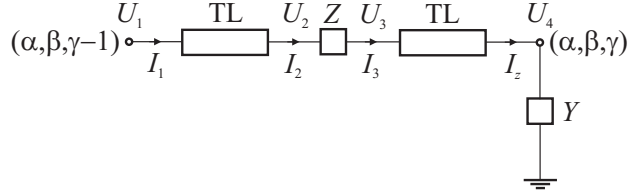, width=0.5\textwidth}
\caption{Part of the 3D structure with transmission lines.}
\label{1D_deriv}
\end{figure}

Next we want to take the transmission lines into account. The
effect of the transmission lines can be derived by first
evaluating a part of the three-dimensional network as the one
shown in Fig.~\ref{1D_deriv} and deriving the relation between the
current that flows into a node and the voltages of adjacent
nodes.

If current $I_1$ is flowing towards node ($\alpha,\beta,\gamma$)
and the current that goes into node ($\alpha,\beta,\gamma$) from
the left is $I_z$, then  using the ABCD-matrix for a transmission
line we get: \e U_1=A_{\rm t}U_2+B_{\rm t}I_2 \l{U_1}, \f \e
I_1=C_{\rm t}U_2+D_{\rm t}I_2 \l{I_1},  \f where \e \left(
\begin{array}{ccc}
A_{\rm t} & B_{\rm t}  \\
C_{\rm t} & D_{\rm t}  \end{array} \right) = \left(
\begin{array}{ccc}
\cos(k_0d/2) & jZ_0 \sin(k_0d/2)  \\
jZ_0^{-1} \sin(k_0d/2) & \cos(k_0d/2)  \end{array} \right).
\l{ABCD} \f $k_0$ in~\r{ABCD} is the wavenumber of waves in the
transmission lines. From~\r{U_1} and~\r{I_1} we can solve $I_1$
and $I_2$ as functions of $U_1$ and $U_2$: \e I_1=\frac{U_2(B_{\rm
t}C_{\rm t}-A_{\rm t}D_{\rm t})+D_{\rm t}U_1}{B_{\rm t}}, \f \e
I_2=\frac{U_1-A_{\rm t}U_2}{B_{\rm t}}. \f Similarly for $I_3$ and
$I_z$ we get: \e I_3=\frac{U_4(B_{\rm t}C_{\rm t}-A_{\rm t}D_{\rm
t})+D_{\rm t}U_3}{B_{\rm t}}, \f \e I_z=\frac{U_3-A_{\rm
t}U_4}{B_{\rm t}}. \l{I_z} \f Next we can derive two equations for
the current flowing through the series impedance $Z$ and solve
$U_2$ from both of them: \e I_2=Z^{-1}(U_2-U_3)=\frac{U_1-A_{\rm
t}U_2}{B_{\rm t}},\f \e \Leftrightarrow U_2=\frac{U_1+Z^{-1}B_{\rm
t}U_3}{A_{\rm t}+Z^{-1}B_{\rm t}}; \f

\e I_3=Z^{-1}(U_2-U_3)=\frac{U_4(B_{\rm t}C_{\rm t}-D_{\rm
t}A_{\rm t})+D_{\rm t}U_3}{B_{\rm t}} \l{I_3}, \f \e
\Leftrightarrow U_2=\frac{U_4(B_{\rm t}C_{\rm t}-D_{\rm t}A_{\rm
t})+U_3(D_{\rm t}+Z^{-1}B_{\rm t})}{Z^{-1}B_{\rm t}}. \l{U_2} \f
If we let $U_2$ in both equations be equal, we can solve $U_3$ as
a function of $U_1$ and $U_4$: \e U_3=\frac{U_1Z^{-1}B_{\rm
t}-U_4(B_{\rm t}C_{\rm t}-D_{\rm t}A_{\rm t})(A_{\rm
t}+Z^{-1}B_{\rm t})}{(D_{\rm t}+Z^{-1}B_{\rm t})(A_{\rm
t}+Z^{-1}B_{\rm t})-Z^{-2}B_{\rm t}^2} . \l{U_3} \f

In order to derive an equation for $I_z$ [the current that flows
into node ($\alpha,\beta,\gamma$) from the direction of node
($\alpha,\beta,\gamma-1$)] as a function of $U_1$ and $U_4$, we
insert~\r{U_3} into~\r{I_z} and get \e I_z=\frac{Z^{-1}B_{\rm
t}U_1 -(B_{\rm t}C_{\rm t}-D_{\rm t}A_{\rm t})(A_{\rm
t}+Z^{-1}B_{\rm t})U_4}{[(D_{\rm t}+Z^{-1}B_{\rm t})(A_{\rm
t}+Z^{-1}B_{\rm t})-Z^{-2}B_{\rm t}^2]B_{\rm t}}-\frac{A_{\rm
t}}{B_{\rm t}}U_4. \f If we use $U_1=U_{\alpha,\beta,\gamma-1}$
and $U_4=U_{\alpha,\beta,\gamma}$, then $I_z=S_{\rm
BW}U_{\alpha,\beta,\gamma-1}+K_{\rm BW}U_{\alpha,\beta,\gamma}$.
Because of the symmetry we can derive the dispersion relation
exactly the same way as in~\r{simple_1}~--~\r{simple_3}, and for
the case $Z=1/j\omega C$, $Y=1/j\omega L$ the result is \e
\cos(q_{x})+\cos(q_{y})+\cos(q_{z})=\frac{1}{2j\omega LS_{\rm
BW}}-3\frac{K_{\rm BW}}{S_{\rm BW}},  \l{disp_rel_BW} \f where \e
S_{\rm BW}=\frac{j\omega C}{(D_{\rm t}+j\omega CB_{\rm t})(A_{\rm
t}+j\omega CB_{\rm t})+\omega^2 C^2B_{\rm t}^2} \l{S_BW}, \f \e
K_{\rm BW}=\frac{-(B_{\rm t}C_{\rm t}-D_{\rm t}A_{\rm t})(A_{\rm
t}+j\omega CB_{\rm t})}{[(D_{\rm t}+j\omega CB_{\rm t})(A_{\rm
t}+j\omega CB_{\rm t})+\omega^{2}C^{2}B_{\rm t}^2]B_{\rm
t}}-\frac{A_{\rm t}}{B_{\rm t}} \l{K_BW} .\f\newline

To derive the dispersion relation for the forward-wave network, we
can use the equations derived for the backward-wave network letting
$C\rightarrow\infty$ and $L\rightarrow\infty$. This way
we get from~\r{S_BW} and~\r{K_BW} the
following equations for $S_{\rm FW}$ and $K_{\rm FW}$:
\e S_{\rm FW}=\frac{1}{B_{\rm t}(A_{\rm t}+D_{\rm t})}, \f \e
K_{\rm FW}=-\frac{B_{\rm t}C_{\rm t}-D_{\rm t}A_{\rm t}}{A_{\rm
t}+D_{\rm t}}\frac{1}{B_{\rm t}}-\frac{A_{\rm t}}{B_{\rm t}}.
\f
From~\r{disp_rel_BW} we get the dispersion relation:
\e \cos(q_{x})+\cos(q_{y})+\cos(q_{z})=-3\frac{K_{\rm FW}}{S_{\rm
FW}}. \l{disp_rel_FW} \f

\subsection{Typical dispersion curves}

\begin{figure}[ht]
\centering \epsfig{file=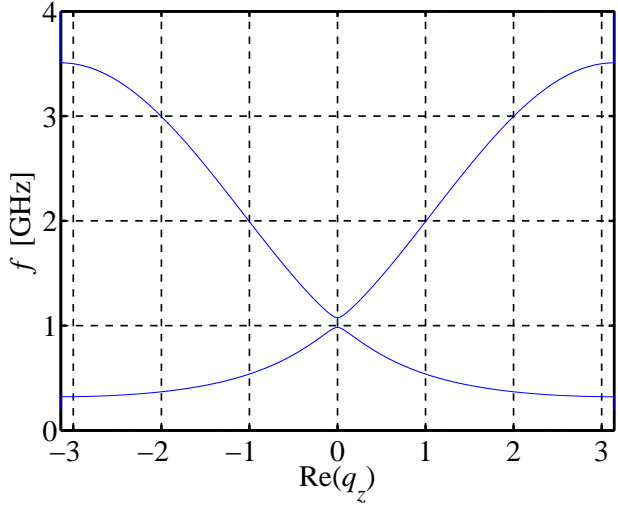, width=0.5\textwidth}
\caption{Dispersion curve for a backward-wave network. $C=5$ pF,
$L=10$ nH, $d=0.012$ m, $\varepsilon_r=2.33$, $Z_{\rm 0,TL}=85$
Ohm.} \label{dispersion_BW}
\end{figure}

\begin{figure}[ht]
\centering \epsfig{file=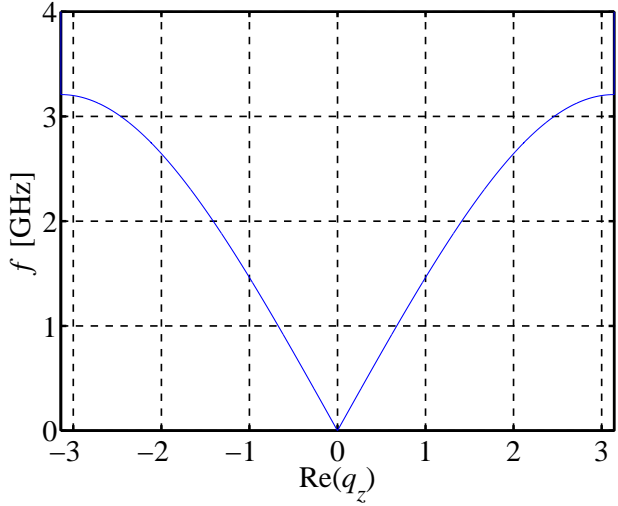, width=0.5\textwidth}
\caption{Dispersion curve for a forward-wave network. $d=0.012$ m,
$\varepsilon_r=2.33$, $Z_{\rm 0,TL}=85$ Ohm.}
\label{dispersion_FW}
\end{figure}

Dispersion curves for backward-wave and forward-wave networks can
be plotted if the values of the transmission line parameters and
\textit{L} and \textit{C} are fixed. Let us choose the parameters
of the TLs and the lumped components as: $L=10$ nH, $C=5$ pF,
$d=0.012$ m (the period of the network), $Z_{\rm 0,TL,BW}=85$ Ohm,
$Z_{\rm 0,TL,FW}=85$ Ohm (characteristic impedances of the TLs).
See Figs.~\ref{dispersion_BW} and~\ref{dispersion_FW} for examples
of dispersion curves when a $z$-directed plane wave is considered
(i.e. $q_{x}=q_{y}=0$). $k_0=\omega \sqrt{\varepsilon_r}/{c}$,
where $c$ is the speed of light in vacuum. Notice that the BW
network supports backward-waves only in the region where 0.32
GHz$\leq f \leq$0.98 GHz. Above that frequency band and the
following stopband, the BW network works as a normal FW network
until the next stopband appears.


\newpage

\begin{figure}[ht]
\centering \epsfig{file=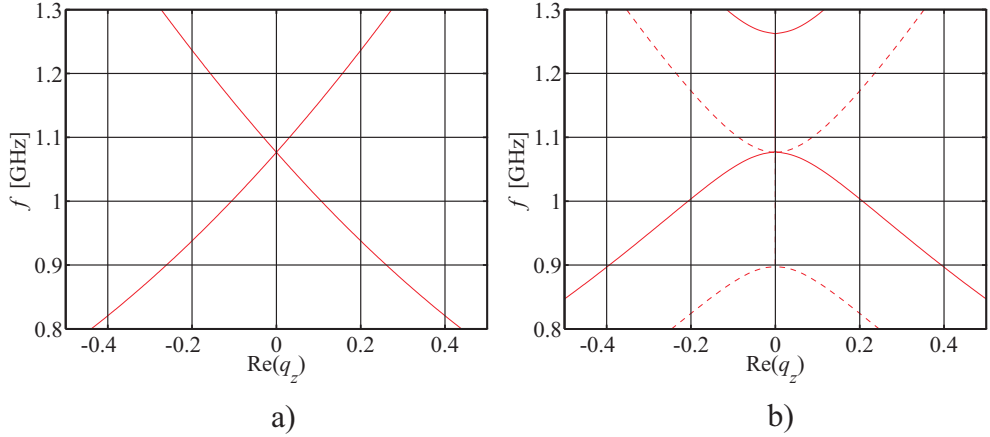, width=0.8\textwidth}
\caption{Dispersion curves for the backward-wave network. $L=10$
nH, $d=0.012$ m, $\varepsilon_r=2.33$, $Z_{\rm 0,TL}=85$ Ohm. a)
$C=4.151$ pF. b) Solid line: $C=3$ pF, dashed line: $C=6$ pF.}
\label{dispersion_Cvar}
\end{figure}

By tuning the capacitance $C$ (or inductance $L$), the stopband
between the BW and FW regions shown in Fig.~\ref{dispersion_BW}
can be closed, see Fig.~\ref{dispersion_Cvar}a, where
$C=C_0=4.151$ pF. As can be seen from Fig.~\ref{dispersion_Cvar}b,
by changing the value of $C$ from this ``balanced'' case, the
stopband is formed either by moving the edge of the FW region up
($C<C_0$) or by moving the edge of the BW region down ($C>C_0$).

\section{The characteristic impedances of backward-wave and forward-wave
networks}

Next the characteristic impedance of the backward-wave network is
derived. If we assume that the interface between the two networks
is in the center of capacitor \textit{C} (see
Fig.~\ref{1D_deriv}, where $Z=1/j\omega C$), then we can define
the characteristic impedance as
\e Z_{\rm 0,BW}=\frac{U_2+U_3}{2I_2}=\frac{U_2+U_3}{2I_3}. \f
First we have to express $U_2$, $U_3$ and $I_3$ (or optionally
$I_2$) as functions of $U_1$ and $U_4$. We can use
equations~\r{I_3}~--~\r{U_3}.
If we insert~\r{U_3}
into~\r{U_2} and~\r{I_3}, we find $U_2$, $U_3$ and $I_3$ as
functions of $U_1$ and $U_4$. Therefore we can present $U_2$,
$U_3$ and $I_3$ simply as: $U_2=a_{\rm BW}U_1+b_{\rm BW}U_4$,
$U_3=c_{\rm BW}U_1+d_{\rm BW}U_4$ and $I_3=e_{\rm BW}U_1+f_{\rm
BW}U_4$.

Because $U_4=e^{-jq_{z}}U_1$ for a wave moving along the
$+z$-direction, the characteristic impedance can be expressed as
\e Z_{\rm 0,BW}=\frac{U_2+U_3}{2I_3}=\frac{a_{\rm BW}+b_{\rm
BW}e^{-jq_{z}}+c_{\rm BW}+d_{\rm BW}e^{-jq_{z}}}{2e_{\rm
BW}+2f_{\rm BW}e^{-jq_{z}}},  \l{Z0_BW} \f where \e a_{\rm
BW}=\frac{D_{\rm t}+j\omega CB_{\rm t}}{(D_{\rm t}+j\omega CB_{\rm
t})(A_{\rm t}+j\omega CB_{\rm t})+\omega^2C^2B_{\rm t}^2}, \f \e
b_{\rm BW}=\frac{B_{\rm t}C_{\rm t}-D_{\rm t}A_{\rm t}}{j\omega
CB_{\rm t}}-\frac{D_{\rm t}+j\omega CB_{\rm t}}{j\omega CB_{\rm
t}}\frac{(B_{\rm t}C_{\rm t}-D_{\rm t}A_{\rm t})(A_{\rm t}+j\omega
CB_{\rm t})}{(D_{\rm t}+j\omega CB_{\rm t})(A_{\rm t}+j\omega
CB_{\rm t})+\omega^2C^2B_{\rm t}^2}, \f \e c_{\rm
BW}=\frac{j\omega CB_{\rm t}}{(D_{\rm t}+j\omega CB_{\rm
t})(A_{\rm t}+j\omega CB_{\rm t})+\omega^2C^2B_{\rm t}^2}, \f \e
d_{\rm BW}=-\frac{(B_{\rm t}C_{\rm t}-D_{\rm t}A_{\rm t})(A_{\rm
t}+j\omega CB_{\rm t})}{(D_{\rm t}+j\omega CB_{\rm t})(A_{\rm
t}+j\omega CB_{\rm t})+\omega^2C^2B_{\rm t}^2}, \f \e e_{\rm
BW}=\frac{D_{\rm t}}{B_{\rm t}}\frac{j\omega CB_{\rm t}}{(D_{\rm
t}+j\omega CB_{\rm t})(A_{\rm t}+j\omega CB_{\rm
t})+\omega^2C^2B_{\rm t}^2}, \f \e f_{\rm BW}=\frac{B_{\rm
t}C_{\rm t}-D_{\rm t}A_{\rm t}}{B_{\rm t}}-\frac{D_{\rm t}}{B_{\rm
t}}\frac{(B_{\rm t}C_{\rm t}-D_{\rm t}A_{\rm t})(A_{\rm t}+j\omega
CB_{\rm t})}{(D_{\rm t}+j\omega CB_{\rm t})(A_{\rm t}+j\omega
CB_{\rm t})+\omega^2C^2B_{\rm t}^2} \l{f_BW}. \f

To derive the characteristic impedance of the forward-wave
network, we can use the equations derived for the backward-wave
network if we insert $C\rightarrow\infty$ in them. If this
condition applies, we get from~\r{Z0_BW}~--~\r{f_BW}:
\e Z_{\rm 0,FW}=\frac{U_2+U_3}{2I_3}=\frac{a_{\rm FW}+b_{\rm
FW}e^{-jq_{z}}+c_{\rm FW}+d_{\rm FW}e^{-jq_{z}}}{2e_{\rm
FW}+2f_{\rm FW}e^{-jq_{z}}} ,  \l{Z0_FW} \f
where
\e a_{\rm FW}=c_{\rm FW}=\frac{1}{A_{\rm t}+D_{\rm t}}, \f
\e b_{\rm FW}=d_{\rm FW}=-\frac{B_{\rm t}C_{\rm t}-D_{\rm t}A_{\rm
t}}{A_{\rm t}+D_{\rm t}}, \f
\e e_{\rm FW}=\frac{D_{\rm t}}{B_{\rm t}}\frac{1}{A_{\rm t}+D_{\rm
t}}, \f
\e f_{\rm FW}=\frac{B_{\rm t}C_{\rm t}-D_{\rm t}A_{\rm t}}{B_{\rm
t}}-\frac{D_{\rm t}}{B_{\rm t}}\frac{B_{\rm t}C_{\rm t}-D_{\rm
t}A_{\rm t}}{A_{\rm t}+D_{\rm t}}. \f

\begin{figure}[ht]
\centering \epsfig{file=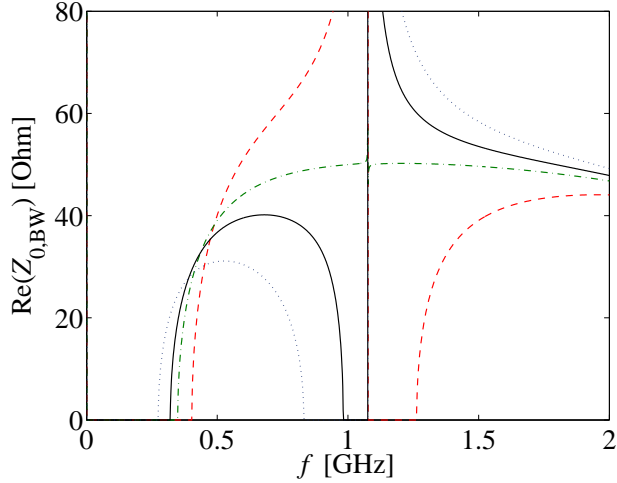,
width=0.5\textwidth} \caption{Characteristic impedance as a
function of frequency for the backward-wave network. $L=10$ nH,
$d=0.012$ m, $\varepsilon_r=2.33$, $Z_{\rm 0,TL}=85$ Ohm. Dashed
line: $C=3$ pF, dash-dotted line: $C=4.151$ pF, solid line: $C=5$
pF, dotted line: $C=7$ pF.} \label{Z0_fvar_BW}
\end{figure}
\begin{figure}[ht]
\centering \epsfig{file=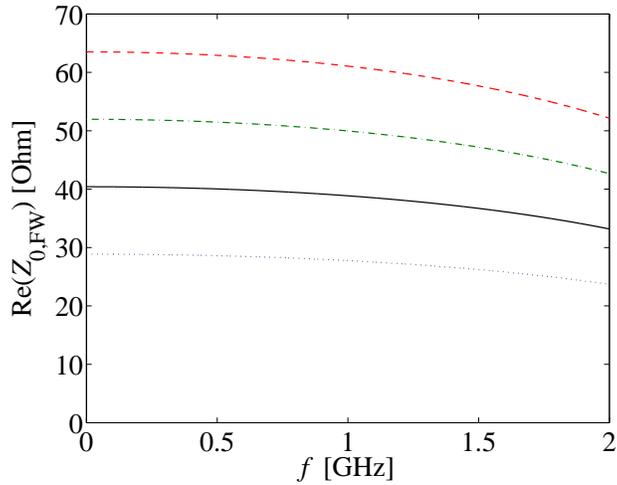,
width=0.5\textwidth} \caption{Characteristic impedance as a
function of frequency for the forward-wave network. $d=0.012$ m,
$\varepsilon_r=2.33$. Dotted line: $Z_{\rm 0,TL}=50$ Ohm, solid
line: $Z_{\rm 0,TL}=70$ Ohm, dash-dotted line: $Z_{\rm 0,TL}=90$
Ohm, dashed line: $Z_{\rm 0,TL}=110$ Ohm.} \label{Z0_fvar_FW}
\end{figure}

$Z_{\rm 0,BW}$ and $Z_{\rm 0,FW}$ can be plotted from~\r{Z0_BW}
and~\r{Z0_FW} as functions of the frequency if the transmission line
parameters and \textit{L} and \textit{C} are fixed. Let us choose
the parameters of the TLs and the lumped components as: $C=5$ pF,
$L=10$ nH, $d=0.012$ m, $\varepsilon_r=2.33$, $Z_{\rm 0,TL,BW}=85$
Ohm, $Z_{\rm 0,TL,FW}=70$ Ohm (characteristic impedances of the
TLs). See Figs.~\ref{Z0_fvar_BW} and~\ref{Z0_fvar_FW} for examples
of the characteristic impedances when a $z$-directed plane wave is
considered (i.e. $q_{x}=q_{y}=0$). The effect of changing $C$ on
the characteristic impedance can be seen in
Fig.~\ref{Z0_fvar_BW}, and the effect of changing $Z_{\rm 0,TL,FW}$
on the characteristic impedance is shown in
Fig.~\ref{Z0_fvar_FW}. Notice that for the BW network the
characteristic impedance is continuous only in the ``balanced''
case ($C=4.151$ pF here), because in the stopbands the real part of the impedance
is zero.

\section{Matching of forward-wave and backward-wave networks}

We consider a perfect lens with axis parallel to the $z$-axis. To
have perfect imaging, the lens should support all spatial
harmonics (i.e. waves with all possible transverse wavenumbers
$k_{\rm t}$) of the source field and for those values of $k_{\rm
t}$, $k_{z,\rm FW}$ and $k_{z,\rm BW}$ should be equal in
magnitude but opposite in sign.

\begin{figure}[ht]
\centering \epsfig{file=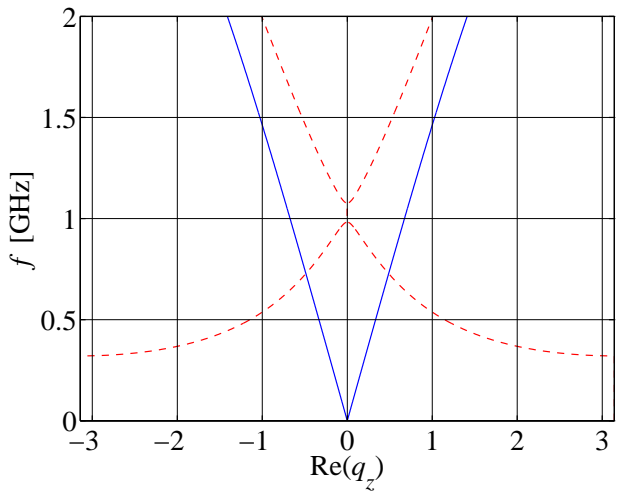,
width=0.5\textwidth} \caption{Dispersion curves for the
forward-wave (solid line) and backward-wave (dashed line)
networks. $C=5$ pF, $L=10$ nH, $d=0.012$ m, $\varepsilon_r=2.33$,
$Z_{\rm 0,TL,BW}=85$ Ohm, $Z_{\rm 0,TL,FW}=70$ Ohm. The transverse
wavenumber $k_{\rm t}=0$.} \label{f_qzvar_BW_and_FW}
\end{figure}

From Fig.~\ref{f_qzvar_BW_and_FW} we can conclude that the
matching of $|k_{z,\rm FW}|$ and $|k_{z,\rm BW}|$ (which
corresponds to a relative refraction index of $-1$) can be achieved
only at one frequency (depending on the parameters of the forward-wave
and backward-wave networks). In Fig.~\ref{f_qzvar_BW_and_FW} this
frequency is $f=0.7279$ GHz. At this frequency the dispersion
curves of the forward-wave and backward-wave networks intersect.
In the analytical form this means that
\e \frac{1}{2j\omega_0 LS_{\rm
BW}(\omega_0)}-3\frac{K_{\rm BW}(\omega_0)}{S_{\rm
BW}(\omega_0)}=-3\frac{K_{\rm FW}(\omega_0)}{S_{\rm FW}(\omega_0)},
\l{disp_ideal} \f
as can be seen from~\r{disp_rel_BW}
and~\r{disp_rel_FW}. The dispersion curves in
Fig.~\ref{f_qzvar_BW_and_FW} are plotted so that $q_{x}$ and
$q_{y}$ are zero (a plane wave moving along the $z$-axis). From
Fig.~\ref{f_qzvar_BW_and_FW} it is seen also that for $f=0.7279$ GHz
$|q_{z,\rm BW}|=|q_{z,\rm FW}|=0.4869$ when $k_{\rm t}=0$. This
means that the absolute value of the maximum wavenumber for
propagating waves is approximately $40.6$ m$^{-1}$.

In addition to matching the wavenumbers (refractive indices), to
realize an ideal ``perfect lens'' the interfaces between the
forward-wave and backward-wave networks should  be also
impedance-matched. If the two regions would not be matched,
reflections from the interface would distort the field patterns
both inside and outside the lens. As can be seen from
Figs.~\ref{Z0_fvar_BW} and~\ref{Z0_fvar_FW}, the characteristic
impedances of the backward-wave network and the forward-wave
network are about 40 Ohms at the frequency where the wavenumbers
are matched (and when $C=5$ pF and $Z_{\rm 0,TL,FW}=70$ Ohm).
Notice that the impedances of the transmission lines in the
forward-wave network have been lowered from 85 Ohm to 70 Ohm to
achieve impedance matching of the forward-wave and backward-wave
networks.

Next the effect of nonzero $k_{\rm t}$ on the matching is
considered (at the optimal frequency). The minimum and maximum
values of $k_{\rm t}$ can be found from $k_{\rm t,min}=-\pi/d$ and
$k_{\rm t,max}=\pi/d$. From~\r{disp_rel_BW} and~\r{disp_rel_FW} we
can plot $k_{z,\rm BW}$ and $k_{z,\rm FW}$ as functions of the
transverse wavenumber ($k_{\rm t}=\sqrt{k_{x}^{2}+k_{y}^{2}}$) if
we fix the frequency. Now $k_{z,\rm BW}$ and $k_{z,\rm FW}$ are
surfaces with variables $k_{x}$ and $k_{y}$. By comparing these
surfaces, it was seen that they are practically the same for all
possible values of $k_{\rm t}$. This happens only at the frequency
$f=0.7279$ GHz, where the dispersion curves of the forward-wave
and backward-wave networks intersect.

\begin{figure}[ht]
\centering \epsfig{file=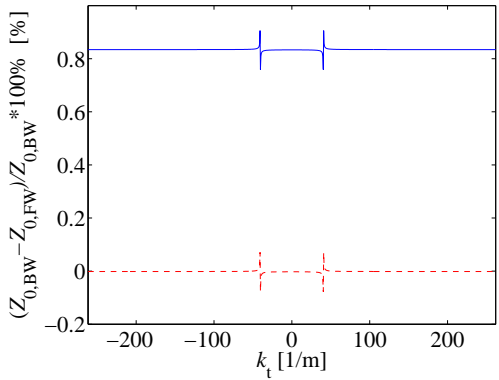, width=0.4\textwidth}
\caption{Relative difference $(Z_{\rm 0,BW}-Z_{\rm 0,FW})/Z_{\rm
0,BW}$ as a function of $k_{\rm t}$. $C=5$ pF, $L=10$ nH,
$d=0.012$ m, $\varepsilon_r=2.33$, $Z_{\rm 0,TL,BW}=85$ Ohm. Solid
line: $Z_{\rm 0,TL,FW}=70$ Ohm, dashed line: $Z_{\rm
0,TL,FW}=70.59$ Ohm. $k_{x}=k_{\rm t}$.} \label{Z0diff_ktvar}
\end{figure}

Because the characteristic impedances are functions of the
ABCD-matrices and $k_{z}$ (and $k_{z}$ is a function of $k_{\rm
t}$), we can plot $Z_{\rm 0,FW}$ and $Z_{\rm 0,BW}$ as functions
of $k_{\rm t}$ if we fix the frequency. Now $Z_{\rm 0,BW}$ and
$Z_{\rm 0,FW}$ are surfaces with variables $k_{x}$ and $k_{y}$. By
comparing these surfaces, it was seen that they are almost the
same (less than one percent difference) for all possible values of
$k_{\rm t}$ at $f=0.7279$ GHz. See Fig.~\ref{Z0diff_ktvar} for a
2D cut of relative difference between such surfaces (now $k_{y}=0$
and therefore $k_{\rm t}=k_{x}$). The ideal ``perfect lens''
situation is achieved with $Z_{\rm 0,TL,FW}=70.59$ Ohm, as can be
seen also from Fig.~\ref{Z0diff_ktvar}.

When the frequency deviates from the optimal value (for
which~\r{disp_ideal} is true), the wavenumbers and characteristic
impedances are no longer matched. The effect of this is distortion
of the image seen in the image plane of the lens.

\newpage

\section{Transmission of the source field through the lens}

\begin{figure}[ht]
\centering \epsfig{file=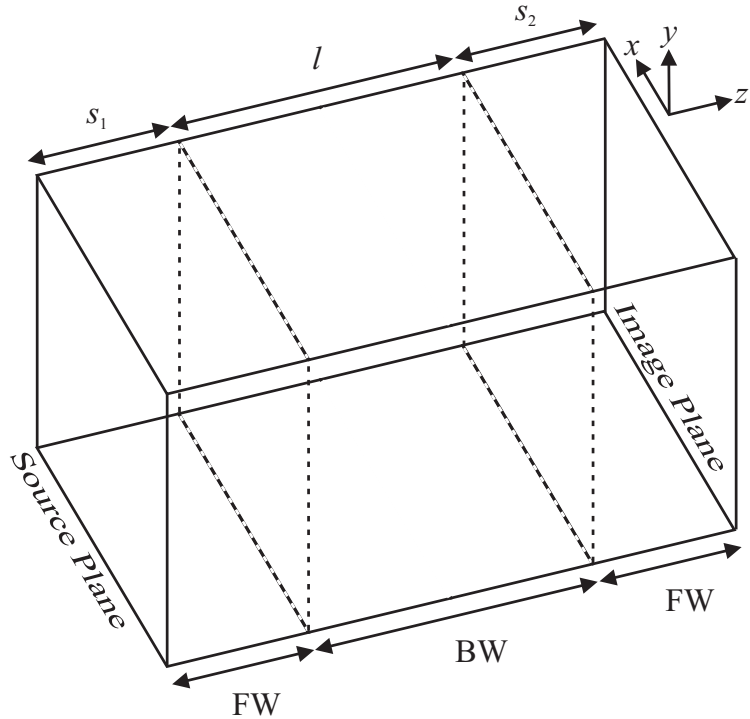, width=0.6\textwidth}
\caption{A 3D perfect lens with distances $l, s_1$ and $s_2$
shown.} \label{perfect_lens}
\end{figure}

The transmission coefficient of the lens can be solved by
considering the incident and reflected fields in the lens system.
Let us assume that the incident field outside the lens has the
unit amplitude, the reflected field outside the lens has amplitude
$R$, the incident field inside the lens has amplitude
$Ae^{-jk_{z,\rm BW} z}$, and the reflected field inside the lens
has amplitude $Be^{+jk_{z,\rm BW} z}$ ($z$ is the distance from
the front edge of the lens). From these values we can form the
following equations (the length of the BW slab is $l$ and the
transmission coefficient of the BW slab is $T_{\rm Lens}$, see
Fig.~\ref{perfect_lens}): \e 1+R=A+B \l{T_1}, \f \e
\frac{1-R}{Z_{\rm 0,FW}}=\frac{A-B}{Z_{\rm 0,BW}} \l{T_2}, \f \e
T_{\rm Lens}=Ae^{-jk_{z,\rm BW} l}+Be^{+jk_{z,\rm BW} l} \l{T_3},
\f \e \frac{T_{\rm Lens}}{Z_{\rm 0,FW}}=\frac{Ae^{-jk_{z,\rm BW}
l}-Be^{+jk_{z,\rm BW} l}}{Z_{\rm 0,BW}}. \f

The resulting equation for the transmission coefficient $T_{\rm
Lens}$ is: \e T_{\rm Lens}(k_{\rm t})=\frac{4Z_{\rm 0,FW}Z_{\rm
0,BW}}{(Z_{\rm 0,FW}+Z_{\rm 0,BW})^{2}e^{+jk_{z,\rm BW}l}-(Z_{\rm
0,FW}-Z_{\rm 0,BW})^{2}e^{-jk_{z,\rm BW}l}}. \l{T}\f The total
transmission from the source plane to the image plane (see
Fig.~\ref{perfect_lens}) is then (distance from source plane to
lens is $s_1$ and distance from lens to image plane is $s_2$) \e
T_{\rm tot}(k_{\rm t})=T_{\rm Lens}(k_{\rm t})e^{-jk_{z,\rm
FW}(s_1+s_2)}. \l{T_tot} \f

The longitudinal wavenumber $k_{z}$ as a function of $k_{\rm t}$
can be found from the dispersion relations. Let us choose $k_{\rm
t}=k_{x}$ and $k_{y}=0$ so we can plot curves instead of surfaces.
$T_{\rm tot}$ as a function of $k_{\rm t}$ can now be plotted if
the frequency is fixed. Let us choose the lengths of the lens
system as the following: $l=0.12$ m, $s_1=0.06$ m, $s_2=0.06$ m.
Now we can choose the frequency at which we want to calculate
$T_{\rm tot}$.

\begin{figure}[ht]
\centering \epsfig{file=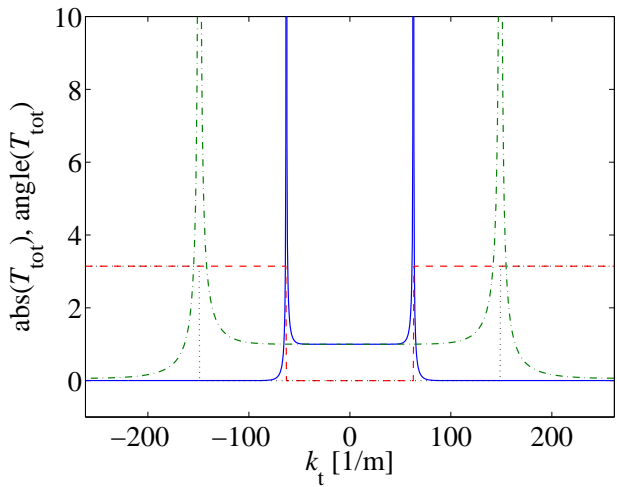, width=0.5\textwidth}
\caption{Total transmission from the source plane to the image
plane as a function of $k_{\rm t}$. $C=5$ pF, $L=10$ nH, $d=0.012$
m, $\varepsilon_r=2.33$, $Z_{\rm 0,TL,BW}=85$ Ohm, $Z_{\rm
0,TL,FW}=70$ Ohm. Case 1: $s_1=s_2=0.06$ m, $l=0.12$ m. Case 2:
$s_1=s_2=0.024$ m, $l=0.048$ m. Solid line: $|T_{\rm tot}|$ (case
1), dashed line: $\arg{(T_{\rm tot})}$ (case 1), dash-dotted line:
$|T_{\rm tot}|$ (case 2), dotted line: $\arg{(T_{\rm tot})}$ (case
2).} \label{T_tot}
\end{figure}

\begin{figure}[ht]
\centering \epsfig{file=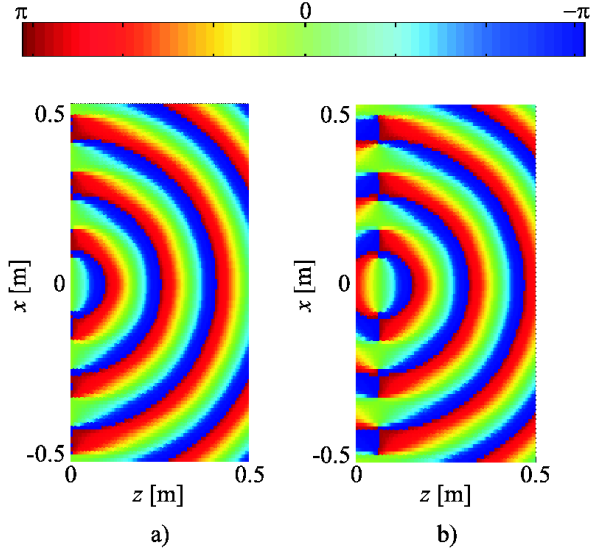, width=0.6\textwidth}
\caption{Phase of electric field (propagating wave). a) Source
plane is at $z=0$ m. b) Back edge of the lens is at $z=0$ m and
image plane is therefore at $z=0.06$ m. $C=5$ pF, $L=10$ nH,
$d=0.012$ m, $\varepsilon_r=2.33$, $Z_{\rm 0,TL,BW}=85$ Ohm,
$Z_{\rm 0,TL,FW}=70$ Ohm. $s_1=s_2=0.06$ m and $l=0.12$ m in
$T_{\rm tot}$.} \label{Source_Image}
\end{figure}

Let us study the transmission properties at the matching frequency
$f=0.7279$ GHz. From~\r{T} and~\r{T_tot} we can plot the magnitude
and phase of $T_{\rm tot}$ as a function of $k_{\rm t}$, see
Fig.~\ref{T_tot}, case 1. From Fig.~\ref{T_tot} it is seen that
the ``lens'' works quite well for the propagating modes ($-40$
m$^{-1}<k_{\rm t}<40$ m$^{-1}$), see Fig.~\ref{Source_Image} for
an example of phase correction in the image plane.

According to Fig.~\ref{T_tot}, for evanescent modes ($k_{\rm
t}<-41$ m$^{-1}, k_{\rm t}>41$ m$^{-1}$)  the ``lens'' works only
in a limited range of $k_{\rm t}$, where the absolute value of the
transmission coefficient $T_{\rm tot}$ is greater than zero. One
can notice that for evanescent modes a mismatch in $k_{z}$ affects
mostly the phase of $ T_{\rm tot} $ in the area of propagating
waves and a mismatch in the characteristic impedances affects
primarily the absolute value of $ T_{\rm tot} $. To improve the
effect of evanescent waves enhancement, the characteristic
impedances should be matched better in the evanescent wave area of
$k_{\rm t}$ (i.e. $k_{\rm t}<-41$ m$^{-1}, k_{\rm t}>41$
m$^{-1}$). There are several ways to achieve a better matching of
the characteristic impedances.

First, there is of course a possibility to change the impedances
of the transmission lines, but this is probably not practically
realizable because it would require very accurate manufacturing
(even a very small deviation from the ideal impedance values
destroys the effect of growing evanescent waves). The tuning of
$Z_{\rm 0,TL,FW}$ was tested and using the exact impedance
required (see Fig.~\ref{Z0diff_ktvar}), the transmission of
evanescent waves was clearly improved. The resonance peaks in
Fig.~\ref{T_tot} were moved further away from the center and the
absolute value of $T_{\rm tot}$ was larger than or equal to unity
approximately for $-100$ m$^{-1}<k_{\rm t}<100$ m$^{-1}$).

Second, there is a possibility to change the frequency and study
if the impedance matching can be made better that way (this also
means that the matching of wavenumbers $k_{z}$ is made worse which
can also destroy the effect of growing evanescent waves). This was
tested and the best results were obtained using frequency
$f=0.72905$ GHz. The region of transmitted $k_{\rm t}$'s was again
increased, i.e. the resonance peaks in Fig.~\ref{T_tot} were moved
further away from the center and the absolute value of $T_{\rm
tot}$ was larger than or equal to unity approximately for $-100$
m$^{-1}<k_{\rm t}<100$ m$^{-1}$).

\begin{figure}[ht]
\centering \epsfig{file=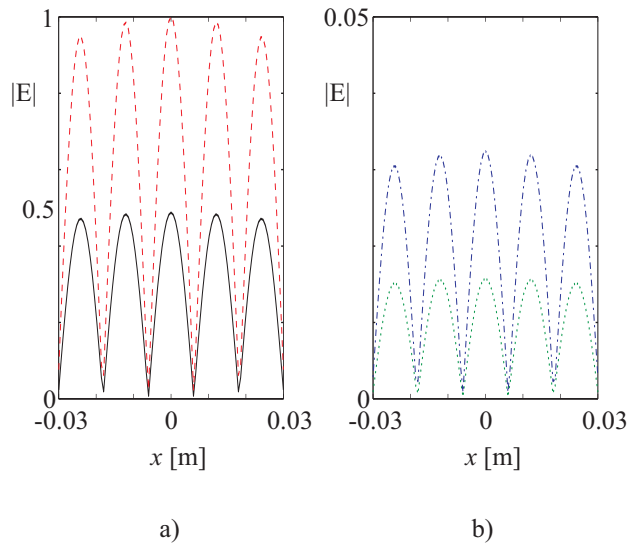,
width=0.5\textwidth} \caption{Absolute value of electric field
(plots normalized to the maximum value) at $f=0.7279$ GHz. a)
Solid line: source field ($z=0$ m), dashed line: field in the back
edge of the lens ($z=0.072$ m). b) Dotted line: field in the front
edge of the lens ($z=0.024$ m), dash-dotted line: field in the
image plane ($z=0.096$ m). The source at $z=0$ produces only
evanescent waves. The source consists of 10 harmonics with $k_{\rm
t}=(2\pi)/(10\cdot 0.012$ m$)\cdot n$, where $n=-5...5$. $C=5$ pF,
$L=10$ nH, $d=0.012$ m, $\varepsilon_r=2.33$, $Z_{\rm 0,TL,BW}=85$
Ohm, $Z_{\rm 0,TL,FW}=70$ Ohm, $s_1=s_2=0.024$ m, $l=0.048$ m.}
\label{source_image_evanescent}
\end{figure}

The third way to enhance the growth of evanescent waves is to
change the length of the ``lens''. From~\r{T} it is seen that the
growth of evanescent waves is destroyed by the term $(Z_{\rm
0,FW}-Z_{\rm 0,BW})^{2}e^{-jk_{z,\rm BW}l}$ in the denominator.
This term can be made smaller by decreasing the length of the
``lens'' $l$. See Fig.~\ref{T_tot}, case 2 ($|T_{\rm tot}|$ is
larger than or equal to unity approximately for $-160$
m$^{-1}<k_{\rm t}<160$ m$^{-1}$) and
Fig.~\ref{source_image_evanescent}, where the distances equal
$l=0.048$ m, $s_1=0.024$ m and $s_2=0.024$ m. From
Fig.~\ref{source_image_evanescent} one can conclude that there is
a significant growth of evanescent waves in the lens.

By using the shortened lens {\it and} at the same time tuning the
frequency appropriately, it was seen that the transmission
coefficient could be made practically ideal (i.e. $|T_{\rm
tot}|=1$ and $\arg{(T_{\rm tot})}=0$ for all possible values of
$k_{\rm t}$). Using the shortened lens (same values as in
Fig.~\ref{source_image_evanescent}) and frequency $f=0.7292$ GHz,
the absolute values of evanescent fields were indeed almost the
same in the image plane and in the source plane (less than one
percent difference).

\section{Suggestions for a practically realizable structure}

\subsection{Proposed structure}

\begin{figure}[ht]
\centering \epsfig{file=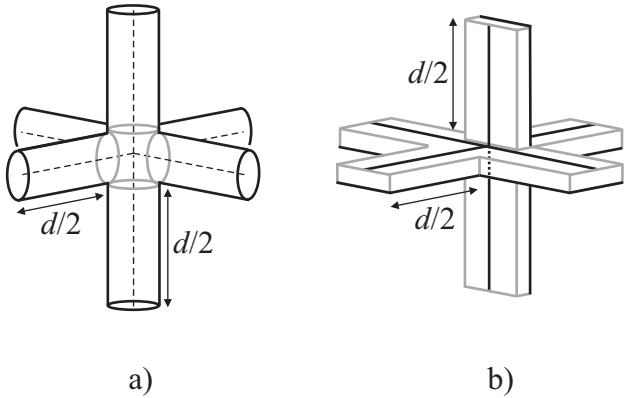,
width=0.5\textwidth} \caption{a) Unit cell of a 3D forward-wave
coaxial network. b) Unit cell of a 3D forward-wave microstrip
network. \textit{L} and \textit{C} can be easily implemented as
lumped components for both types of backward-wave networks.}
\label{3D_coax_and_mstrip_unit_cell}
\end{figure}

How to manufacture three-dimensional transmission line networks?
The main problem is the ground plane, which should exist in all
three dimensions. One solution would be to use coaxial
transmission lines (regular in the forward-wave network and loaded
with lumped \textit{L}- and \textit{C}-components in the
backward-wave network) as shown in
Fig.~\ref{3D_coax_and_mstrip_unit_cell}a. This structure is
realizable, but we propose a simpler structure based on microstrip
lines, as presented in Fig.~\ref{3D_coax_and_mstrip_unit_cell}b.
The problem with microstrip lines is of course the design of
intersections where the transmission lines from six directions
meet. This problem can be overcome by using ground planes which
have holes in them at these intersection points. This way the
conducting strip can be taken through the substrate and thus
connection of the vertical conducting strips becomes possible.

\subsection{Simulation results}

The proposed structure (see
Fig.~\ref{3D_coax_and_mstrip_unit_cell}b) has been simulated in
Ansoft HFSS (Version 9.2.1). Due to complexity of the structure
and the limited calculation power available, the three-dimensional
structure was simulated only near the first lens interface of
Fig.~\ref{perfect_lens}. The simulated model had $10\times 3\times
3$ ($x\times y\times z$) unit cells in the forward-wave region and
$10\times 3\times 3$ unit cells in the backward-wave region. The
properties of the transmission lines and lumped elements were the
same as in Fig.~\ref{f_qzvar_BW_and_FW}. The edges of the system
were terminated with matched loads to prevent reflections ($R_{\rm
BW}=85$ Ohm in the backward-wave region and $R_{\rm FW}=70$ Ohm in
the forward-wave region). Different types of source fields (plane
waves with different incidence angles and a point source) were
tested and in all cases negative refraction was observed at the
interface between the forward-wave and backward-wave networks at
the expected frequency ($f=0.7279$ GHz).

\begin{figure}[ht]
\centering \epsfig{file=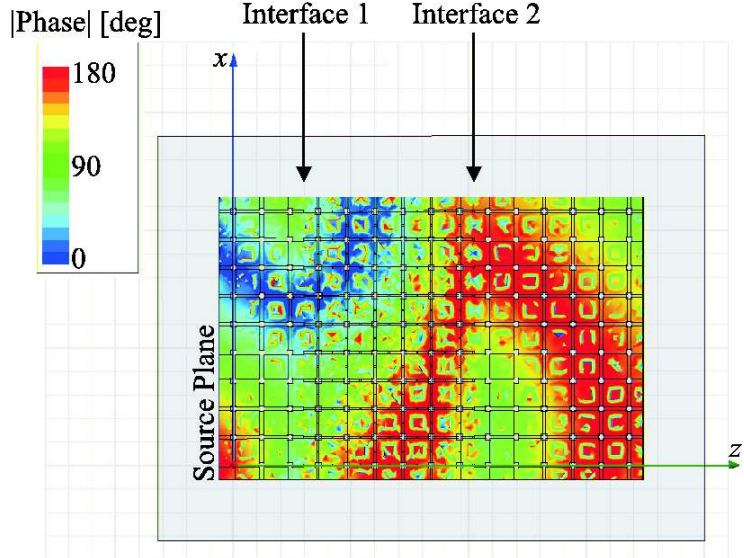, width=0.6\textwidth}
\caption{Plot of $\arg (E_z)$ in a two-dimensional part of the
proposed lens system. The source field is a plane wave with
incidence angle $\theta=25^{0}$. The source field is excited at
the left edge of the system. $C=5$ pF, $L=10$ nH, $d=0.012$ m,
$\varepsilon_r=2.33$, $Z_{\rm 0,TL,BW}=85$ Ohm, $Z_{\rm
0,TL,FW}=70$ Ohm.} \label{2Dplot}
\end{figure}

A two-dimensional cut of the proposed structure was simulated as a
complete ``lens'' system. Again negative refraction was seen at
both interfaces, and therefore also focusing of propagating waves
was observed. See Fig.~\ref{2Dplot} for the plot of the phase of
the electric field in the two-dimensional simulation. The source
field is excited at the left edge of the system in
Fig.~\ref{2Dplot}. When the field magnitude $|E|$ is plotted and
animated as a function of phase, it is clearly seen that the phase
propagates to the right in the forward-wave regions and to the
left in the backward-wave region. The energy of course propagates
to the right in all regions.

\newpage

\section{Conclusions}

In this paper we have introduced and studied a three-dimensional
transmission-line network which is a circuit analogy of the
superlens proposed by Pendry. The structure is a 3D-network of
interconnected loaded transmission lines. Choosing appropriate
loads we realize forward-wave (FW) and backward-wave (BW) regions
in the network. The dispersion equations and analytical
expressions for the characteristic impedances for waves in FW and
BW regions have been derived. A special attention has been given
to the problem of impedance and refraction index matching of FW
and BW regions. From the derived dispersion equations it has been
seen that there exist such a frequency at which the corresponding
isofrequency surfaces for FW and BW regions coincide.
Theoretically this can provide distortion-less focusing of the
propagating modes {\em if} the wave impedances of FW and BW
regions are also well matched. Impedance matching becomes even
more important when the evanescent modes are taken into account.
In this paper we have shown that the wave impedances can be
matched at least within 1\% accuracy or better if the
characteristic impedances of the transmission lines are properly
tuned. However, from the practical point of view an accuracy
better than 1\% becomes hardly realizable. It has been shown that
decreasing the thickness of the BW region reduces the negative
effect of the impedance mismatch, while the amplification of the
evanescent modes is preserved. We have also outlined a couple of
prospective designs of the perfect lens discussed in this paper
and numerically simulated their performance.

\subsection*{Acknowledgment}
This work has been done within the frame of the
\textit{Metamorphose} Network of Excellence and partially funded
by the Academy of Finland and TEKES through the
Center-of-Excellence program. The authors would like to thank Dr.
Mikhail Lapine for bringing paper \cite{Grbic4} to their attention
and for helpful discussions.

\end{document}